\documentclass{article}
\usepackage{epsfig}

\def\Rad#1{%
	\begingroup
	\def\RadTempCs{{#1}}\let\RdxTempCs=\empty
}

\def\DoRad{%
	\relax
	\ifx \RdxTempCs\empty
		\sqrt\RadTempCs
	\else
		\root \RdxTempCs \of \RadTempCs
	\fi
	\endgroup
}

\def\MthAcnt#1#2{#2{#1}}

\begin{document}

\title{Controlling Smart Matter}
\author{Tad Hogg and Bernardo A. Huberman \\
	Xerox Palo Alto Research Center \\
	Palo Alto, CA 94304}

\maketitle

\begin{abstract}
Smart matter consists of many sensors, computers and actuators embedded
within materials. These microelectromechanical systems allow 
properties of the materials to be adjusted under program control. 
In this context, we study the behavior of several
organizations for distributed control of unstable physical systems and
show how a hierarchical organization is a reasonable compromise between
rapid local responses with simple communication and the use of global
knowledge. Using the properties of random matrices, we show that this 
holds not only in ideal situations but also when
imperfections and delays are present in the system. We also introduce a
new control organization, the multihierarchy, and show it is better than
a hierarchy in achieving stability. The multihierarchy also has a
position invariant response that can control disturbances at the
appropriate scale and location. 
\end{abstract}

\section{ Introduction}
 Embedding microscopic sensors, computers and actuators into
materials allows physical systems to actively monitor and respond to
their environments in precisely controlled ways. This is particularly so
for microelectromechanical systems (MEMS)~\cite{bryzek94,berlin95,web.mems96} 
where the devices are
fabricated together in single silicon wafers. Applications include
environmental monitors, drag reduction in fluid flow, compact data
storage and improved material properties. 

 In many applications of such systems the relevant mechanical
processes are slow compared to sensor, computation and communication
speeds. These cases give rise to a {\em smart matter}
regime, where control programs execute many steps within the time
available for responding to mechanical changes. In addition to the
challenge of manufacturing such systems, a key difficulty in realizing
the potential of smart matter is developing appropriate control
programs. This is due to the need to robustly coordinate a physically
distributed real-time response with many elements in the face of
failures, delays, an unpredictable environment and a limited ability to
accurately model the system{'}s behavior. When the system contains many
elements, these characteristics limit the effectiveness of conventional
control algorithms, which rely on a single global processor with rapid
access to the full state of the system and detailed knowledge of its
behavior.  

 A more robust approach for such systems uses a collection of
distributed autonomous processes, or {\em agents}, that
each deal with a limited part of the overall control problem. Individual
agents can be associated with each sensor or actuator in the material,
or with various aggregations of these devices, to provide a mapping
between agents and physical location. This leads to a community of
computational agents which, in their interactions, strategies, and
competition for resources, resemble natural ecosystems~\cite{huberman88}. 
Distributed controls allow the system as
a whole to adapt to changes in the environment or disturbances to
individual components~\cite{hogg91a}. One
disadvantage however, is that multiagent system behavior is much more
difficult to characterize formally, thus necessitating a more
phenomenological approach, as used in this paper. 

 Multiagent systems have been extensively studied in the context of
distributed problem solving~\cite{durfee91,gasser89,lesser95,glance95a}. 
They have also
been applied to problems involved in acting in the physical world, such
as distributed traffic control~\cite{nagel94},
flexible manufacturing~\cite{upton92}, the design
of robotic systems~\cite{sanderson83}, and
self-assembly of structures~\cite{semela95}.
However, the use of multiagent systems for controlling smart matter is a
challenging new application due to the very tight coupling between the
computational agents and their embedding in physical space.
Specifically, in addition to computational interactions between agents
from the exchange of information, there are mechanical interactions
whose strength decreases with the physical distance between them. 

 An important issue in designing multiagent systems to control smart
matter is how the agents are to be organized. From the point of view of
an individual agent, the organization determines what other agents it
communicates with to receive information about the state of the system
for use in its control computation. This computational organization acts
in conjunction with the physical couplings among the agents based on
their physical locations within the material. 

 In this paper we investigate the organization of multiagent systems
for controlling smart matter. Specifically, we focus on a particular
application of smart matter: maintaining a configuration that is
unstable in the absence of controls. This situation has been studied in
the case of small systems with global controls~\cite{berlin94}. 
The organizations we study are:
\begin{enumerate}
\item A {\em global} organization, in which complete
information is immediately available to agents when they make control
decisions. Although not feasible in practice for many large scale
systems, this provides an ideal behavior to compare with that of other
organizations.
\item A {\em local} organization, in which each agent
communicates with spatially close neighbors.
\item A simple {\em hierarchy}, in which groups of
spatially adjacent agents are considered in aggregate, with manager
agents handling the information appropriate to the aggregated region.
This allows control decisions to be made at various scales while
maintaining local responsiveness and limiting the required
communication.
\item A {\em multihierarchy}, which consists of a
collection of overlapping hierarchies, in which each agent is
simultaneously a manager at each level of aggregation. This novel
architecture has been studied in the context of image recognition
systems~\cite{mahoney95}.
\end{enumerate}

 To make this discussion concrete, we use a simple example of smart
matter consisting of an unstable elastic chain of elements with nearest
neighbor interactions and sensors and controllers that act on each
element. We show how a hierarchy is a reasonable compromise between
rapid local responses with simple communication and the use of global
knowledge. This holds not only in ideal situations but also when
imperfections and delays are present in the system. We also show how a
multihierarchy control organization is somewhat better than a hierarchy
in achieving stability. In addition, the position invariant response of
the multihierarchy can control disturbances at an appropriate scale and
location. 

 This study allows us to determine how well these organizations use
the available control force to achieve stability, both when the agents
have a good model of the system{'}s behavior and when they operate with
imperfect models or with delays in the actuator response. Finally, we
investigate the conditions under which the large transient response of
the system can become explosive even if the agents maintain
stability.

\section{ Control Dynamics for Smart Matter}
 In this paper we consider the problem of maintaining a spatially
extended physical system near an unstable configuration. We suppose that
sensors and actuators are embedded in this system at various locations.
Associated with these devices are computational agents that use the
sensor information to determine appropriate actuator forces. The overall
system dynamics can be viewed as a combination of the behavior at the
location of these agents and the behavior of the material between the
agent locations. For the latter, the dynamics consists of high frequency
oscillations that, as we will see below, are not important for the
overall stability. This is because stability is primarily determined by
the behavior of the lowest frequency modes. We assume that there are
enough agents so that their typical spacing is much smaller than the
wavelengths associated with these lowest modes. Hence, to focus on the
lower frequency dynamics it is sufficient to characterize the system by
the displacements at the locations of the agents only. In this case, the
high-frequency dynamics of the physical substrate between agents does
not significantly affect overall stability. Instead, the substrate
serves only to couple the agents{'} displacements. Thus, for our purposes,
the system can be described by a vector \(
{\bf u}{\left( t\right) }\) giving the displacement of each agent at time
{\it t}, and the corresponding velocities
\(
{\bf \MthAcnt {{\bf u}}{\dot }}{\left( t\right) }
\). 

 In general, the relevant physical forces will depend on these
displacement and velocity vectors. If the control mechanism is
successful, the departures from the desired configuration will remain
small so the forces will be dominated by the linear response of the
system. We thus consider a linear system where the forces depend
linearly on displacements and velocities. The dynamical equation
describing the system will then be given by:
\begin{equation}\label{x:dynamics}\vcenter{\halign{\strut\hfil#\hfil&#\hfil\cr 
$\displaystyle{{{d^{2}{\bf u}}\over{dt^{2}}}
=A{\bf u}-G\MthAcnt {{\bf u}}{\dot }}$\cr 
}}\end{equation}
The coupling matrices {\it A} and
{\it G} determine the forces produced by
the displacements and velocities, respectively. Typically, the
velocity-dependent force arises from damping in the physical system. For
simplicity, we assume that the damping is the same for all parts of the
system. Specifically, we take {\it G} to be
a diagonal matrix all of whose elements are equal to the same positive
value, which we denote as $\Gamma$, i.e., \(
G=\Gamma I\) where {\it I} is the
identity matrix. We also assume that the matrix
{\it A} is such as to make the uncontrolled
system unstable, i.e., most small initial displacements eventually
become arbitrarily large. 

 A number of control methods can be used. We focus on a simple case
where the controls act only on the current state of the physical system
rather than, for example, on averages of past behavior or directly in
response to arbitrary messages from other agents. In this case, the
control programs are particularly simple in that they need not save
information about the system{'}s behavior, and the control force becomes a
function only of the agent displacements and velocities as provided by
the sensors. By considering the behavior near the desired configuration,
we can linearize the control function. For the part of the control
depending on the displacements the force becomes \(
-C{\bf u}\), where the elements of the control matrix
{\it C} are determined by the particular
organization of sensors and actuators, their communication architecture
and the control algorithm. 

 There could also be a velocity-dependent control force. This would
amount to an additional damping on the system. Some amount of damping is
important because otherwise a successful control could keep any original
perturbation from growing, but not necessarily from oscillating. As
additional perturbations are added, the total amplitude will tend to
grow. While the linear system will continue to maintain control, the
ever larger oscillations will require larger control forces so nonlinear
behavior will become important and the actuators will eventually fail.
Thus it is important to damp out the oscillations in addition to
preventing their growth. This damping could be part of the physical
behavior of the uncontrolled system, or added as part of the control
mechanism. Since the latter case amounts to an addition to the physical
damping, we do not consider it explicitly here but rather assume that
\(
\Gamma {\ifmmode>\else$>$\fi}0\) so there is some damping in the system. 

 With this control mechanism, the dynamics of
Eq.~\ref{x:dynamics} becomes
\begin{equation}\label{x:matrix dynamics}\vcenter{\halign{\strut\hfil#\hfil&#\hfil\cr 
$\displaystyle{{{d^{2}{\bf u}}\over{dt^{2}}}
=M{\bf u}-G\MthAcnt {{\bf u}}{\dot }}$\cr 
}}\end{equation}
where the new matrix {\it M} is given by
\begin{equation}\vcenter{\halign{\strut\hfil#\hfil&#\hfil\cr 
$\displaystyle{M=A-C}$\cr 
}}\end{equation}
The solution of Eq.~\ref{x:matrix dynamics} can
be written as
\begin{equation}\label{x:solution}\vcenter{\halign{\strut\hfil#\hfil&#\hfil\cr 
$\displaystyle{{\bf u}{\left( t\right) }=\sum 
_{k}{\bf e}^{{\left( k\right) }
}{\left( a^{+}_{k}e^{\Lambda ^{%
+}_{k}t}+a^{-}_{k}e^{\Lambda ^{%
-}_{k}t}\right) }}$\cr 
}}\end{equation}
where \(
{\bf e}^{{\left( k\right) }}\) is the k${}^{th}$ normalized eigenvector of
the matrix {\it M}, with eigenvalue
\(
\lambda _{k}\), i.e., \(
M{\bf e}^{{\left( k\right) }}=\lambda _{%
k}{\bf e}^{{\left( k\right) }}
\). Substituting Eq.~\ref{x:solution} into Eq.~\ref{x:matrix dynamics}
gives a condition that \(
\Lambda ^{{\ifmmode\pm\else$\pm$\fi}}_{k}\) must satisfy, namely \(
\Lambda ^{2}_{k}+\Gamma \Lambda _{k}
-\lambda _{k}=0\) so that
\begin{equation}\label{x:eigenvalues}\vcenter{\halign{\strut\hfil#\hfil&#\hfil\cr 
$\displaystyle{\Lambda ^{{\ifmmode\pm\else$\pm$\fi}}_{k}=-{{\Gamma 
}\over{2}}{\ifmmode\pm\else$\pm$\fi}\Rad{{{\Gamma ^{2}
}\over{4}}+\lambda _{k}}\DoRad }$\cr 
}}\end{equation}
The coefficients \(
a^{{\ifmmode\pm\else$\pm$\fi}}_{k}\) in Eq.~\ref{x:solution} are
determined from the initial conditions, i.e., the displacement and
velocity of each element in the chain at time 0. 

 A necessary property of a successful control mechanism is that it
achieve stability, i.e., prevent any solution from growing arbitrarily
large as time increases. From Eq.~\ref{x:solution}, we see this is 
equivalent to the requirement that
\(
\hbox{\rm Re}{\left( \Lambda ^{{\ifmmode\pm\else$\pm$\fi}}_{k}
\right) }{\ifmmode\leq\else$\leq$\fi}0\) for all {\it k}. From
Eq.~\ref{x:eigenvalues}, \(
\hbox{\rm Re}{\left( \Lambda ^{-}_{k}\right) }
{\ifmmode\leq\else$\leq$\fi}0\) is always true, but \(
\hbox{\rm Re}{\left( \Lambda ^{+}_{k}\right) }
{\ifmmode\leq\else$\leq$\fi}0\) requires
\begin{equation}\label{x:criterion}\vcenter{\halign{\strut\hfil#\hfil&#\hfil\cr 
$\displaystyle{\lambda _{k}{\ifmmode\leq\else$\leq$\fi}0}$\cr 
}}\end{equation}
for all {\it k}. This thus gives a
stability criterion for a control mechanism: no eigenvalue of the
resulting {\it M} matrix can be positive.
This criterion is independent of the amount of damping in the system,
and is similar to standard stability criteria for distributed
systems~\cite{wang73}. Conversely, this discussion
also gives a criterion for the uncontrolled system to be unstable,
namely the largest eigenvalue of {\it A} is
positive.

\section{ Control Organizations}{\it }
 In this section we describe a number of ways to organize
distributed control. An organization determines the information about
the system available to each agent for its control computation. The
effectiveness of these organizations is evaluated in the remaining
sections of this paper. 

\subsection{ Global}
 In the global organization, each agent has full access to the
positions of all the others. In many respects, this is the simplest
organization to program since the desired global behaviors can be
controlled directly. It is commonly used for controlling systems with a
few agents but does not scale well to larger numbers due to the
communication requirements of the sensors and actuators and requirements
for an accurate physical model.

\subsection{ Local}
 The simplest organization to implement physically is one in which
each agent decides what to do based only on its own state and not on
that of others. While this architecture has the advantage of requiring
no communication or coordination among the various controllers, the
effects of changes in one control element are communicated to others
only through physical interactions, thus limiting the ability of agents
to {``}plan ahead{''} or respond rapidly to problems in distant
regions of the system. 

 A direct extension of this organization allows information from
nearest neighbors to be used as well. This is relatively easy to
implement since the communication pattern fits readily within the
physical layout of the agents.

\subsection{ Hierarchy}
 A common compromise between the use of all the information provided
by the global organization and the simple construction of the local
organization is a hierarchically organized system. In this case, the
control system is divided into a number of levels, with
{``}managers{''} at each level responsible for communicating with
subordinate agents below and superior agents above. In terms of
information flow, the managers can aggregate the states of lower level
agents and make this aggregate information available to agents in other
parts of the system. In this way, each agent has, in addition to its own
state, some averaged information about the state on larger scales.
Hierarchical organizations for control have been used in other contexts,
such as controlling vibrations in a stable system~\cite{hall91,how92}. 

 While not as detailed as full global information, a hierarchical
organization provides some contextual information with only a modest
communication overhead. This organization can be viewed in two ways,
giving rise to different types of implementations. First, the managers
can be viewed as actual agents responsible for communicating aggregated
information but not directly connected to sensors or actuators. Because
the managers operate with information averaged over space and time, they
could work more slowly, and perhaps do more complex computations, than
the lowest level agent. This requires implementing two kinds of agents,
with at least somewhat different hardware or software. In another view,
the managers simply represent the behavior of communication channels
between the actual agents. This simpler perspective is more appropriate
for the method of averaging information we discuss since the managers do
not have more complex programs than the lowest level agents. 

 This type of organization introduces an additional distance measure
between two agents: the number of levels in the hierarchy that must be
climbed to find their first common ancestor. This ultrametric
distance~\cite{hartigan67} determines the
aggregation level of the information received, and for some pairs of
agents it differs considerably from their physical separation. 

\begin{figure}
\hspace*{\fill}
\epsfig{file=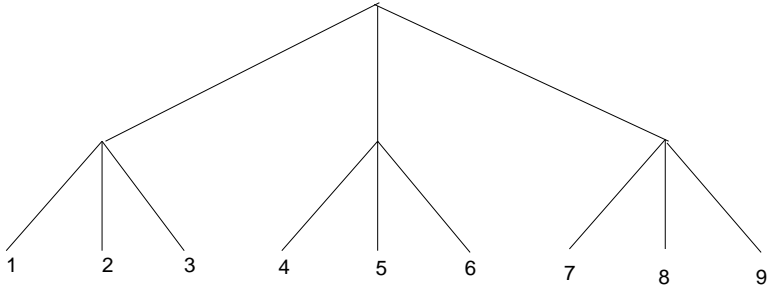,width=4in}
\hspace*{\fill}
\caption{\label{x:tree.ps}An example of a hierarchical organization, 
depicting the grouping of 9 agents into 3 levels.}
\end{figure}

 A simple example of this organization is illustrated in
Fig.~\ref{x:tree.ps}. It depicts the grouping of
nine agents into a hierarchy. Consider agent number 3, with displacement
information \(
u_{3}\). It receives information from its immediate manager,
who conveys aggregate information from agents 1, 2 and 3. From the next
level up,  it receives aggregate information from 1 through 9. 

 More generally, consider a uniform tree structure with branching
ratio {\it b} and depth
{\it d} so that the total number of nodes,
{\it N}, is given by \(
N=b^{d}\). We number the levels of the tree from 0 for the root
at the top down to the leaves at depth
{\it d}. At level
{\it j}, there are \(
b^{j}\) nodes, which we denote by numbers from 1 to
\(
b^{j}\). At level {\it j}, node
{\it x} has parent \(
{\left\lfloor {\left( x-1\right) }/b\right\rfloor }
+1\) at level \(
j-1\) and the set of children \(
{\left\lbrace b{\left( x-1\right) }+i;i=1,\ldots 
,b\right\rbrace }\) at level \(
j+1\). This node has descendents \(
{\left\lbrace b^{d-j}{\left( x-1\right) }
+i;i=1,\ldots ,b^{d-j}\right\rbrace }\) at the bottom of the tree, 
corresponding to the
actual control agents. Alternatively, agent
{\it m}, at the bottom of the tree, has
ancestor \(
x={\left\lfloor {\left( m-1\right) }/b^{d
-j}\right\rfloor }+1\) at level {\it j}, all of
whose descendents at the bottom of the tree have an ultrametric distance
to {\it m} less than or equal to
\(
d-j\). We define 
\begin{equation}\label{x:eta(m,j)}\vcenter{\halign{\strut\hfil#\hfil&#\hfil\cr 
$\displaystyle{\eta {\left( m,j\right) }=b^{d-j}
{\left\lfloor {\left( m-1\right) }/b^{d-j
}\right\rfloor }+1}$\cr 
}}\end{equation}
as the number of the first agent along the bottom of the tree with the
same ancestor at level {\it j} as agent
{\it m}. 

 In this discussion, the focus is on one-dimensional structures with
uniform branching. However, the hierarchy can apply also to 2 or
3-dimensional physical systems where higher levels correspond to larger
physical regions, and nonuniform branching.

\subsection{\label{x:multihierarchy} Multihierarchy}
 One potential difficulty with the hierarchical organization is the
occasional extreme mismatch between the physical and organizational
distances between agents. In particular, agents near an organizational
boundary of a high level part of the hierarchy require many levels in
the hierarchy to communicate with some of their physical neighbors. This
introduces an inhomogeneity in the response of the system in that some
medium-scale perturbations can be controlled entirely within a single
part of the hierarchy while others, crossing these high level
boundaries, require additional levels of communication and aggregation
of information. This means that the hierarchy will not always allow a
response at the most appropriate scale. 

\begin{figure}
\hspace*{\fill}
\epsfig{file=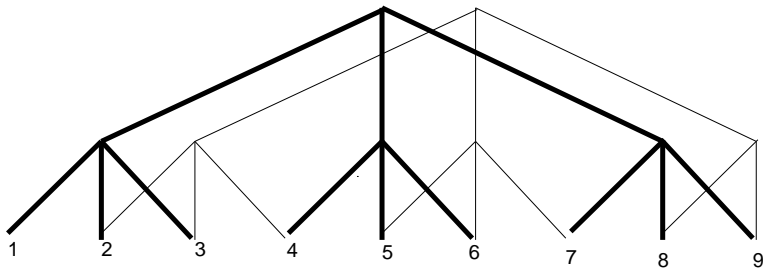,width=4in}
\hspace*{\fill}
\caption{\label{x:multitree.ps}An example of a multihierarchical organization. 
Only two parts are shown, those centered on agents 5 and 6. 
The full organization has a separate hierarchy centered around every agent.}
\end{figure}

 To address this issue, and also provide a more uniform structure
for the agents, we consider the multihierarchy organization that has
been used in the context of machine vision algorithms~\cite{mahoney95}. 
In this case, each agent acts as a
manager at all levels and the organization consists of a set of
interleaved hierarchies. Thus each agent has information from multiple
parents. An example is shown in Fig.~\ref{x:multitree.ps}. Note that 
each agent has three parents, only two of
which are shown in the figure. 

 To complete the specification of the organization, some rule for
combining the information from the multiple parents must be employed.
Simple functions for such rules are the maximum or average of
information received from the parents. Different choices amount to a
focus on different aspects of the problem. For instance, using the
maximum causes the agents to respond to the worst perturbation at any
scale. 

 For a physical system where influences tend to decrease with
physical distance, we use the multihierarchy to provide each agent with
aggregated information centered on its location. Thus each agent uses
information from only one parent, in effect repeating the aggregation
for the hierarchical case but allowing each agent to be in the
{``}center{''} of the hierarchy. That is, each agent gets a
somewhat different average based on distance from that agent, rather
than grouping all agents under a single manager in the hierarchical
case.  

The agents within ultrametric distance
$k$ of agent $m$ are \(
{\left\lbrace m+i;i=-{\left\lfloor b^{k}/2\right\rfloor }
,\ldots ,{\left\lfloor b^{k}/2\right\rfloor }\right\rbrace }
\), but excluding any beyond the edges of the chain,
i.e., we require \(
0{\ifmmode<\else$<$\fi}m+i{\ifmmode\leq\else$\leq$\fi}N\). 
Because of this symmetric form, the distance from
{\it m} to
{\it n} is the same as from
{\it n} to
{\it m}. Let 
\begin{equation}\label{x:eta(min/max)}\vcenter{\halign{\strut\hfil#\hfil&#\hfil\cr 
$\displaystyle{\eta ^{{\left( {\rm min}\right) }
}_{mk}=\max {\left( 1,m-{\left\lfloor 
b^{k}/2\right\rfloor }\right) }}$\cr 
$\displaystyle{\eta ^{{\left( {\rm max}\right) }
}_{mk}=\min {\left( N,m+{\left\lfloor 
b^{k}/2\right\rfloor }\right) }}$\cr 
}}\end{equation}
so all agents \(
\eta ^{{\left( {\rm min}\right) }}_{%
mk}{\ifmmode\leq\else$\leq$\fi}i{\ifmmode\leq\else$\leq$\fi}\eta ^{{\left( {\rm max}\right) }
}_{mk}\) have ultrametric distance of at most
{\it k} from agent
{\it m}. With this definition the maximum
distance can be more than the depth of the tree that would be used for
the single hierarchy. For example, in Fig.~\ref{x:multitree.ps}, agents 
1 and 6 have ultrametric distance 3. The
number of agents within ultrametric distance
{\it k} is then 
\begin{equation}\label{x:Nmk}\vcenter{\halign{\strut\hfil#\hfil&#\hfil\cr 
$\displaystyle{N_{mk}=\eta ^{{\left( max\right) }
}_{mk}-\eta ^{{\left( min\right) }
}_{mk}+1}$\cr 
}}\end{equation}

\section{ Example: An Unstable Elastic Array}
 In what follows we illustrate these ideas by controlling a discrete
system motivated by the classical problem of the buckling instability of
a beam~\cite{naschie90}. Basically the control
problem consists in applying forces perpendicular to the beam so as to
prevent it from buckling when loaded beyond the critical stress. In this
case the unbuckled state corresponds to an unstable fixed point, posing
a challenging control problem. 

 Consider an array of {\it N} elements
of unit mass separated by a distance
{\it a}, with nearest neighbors coupled by
springs with force constants \(
\alpha \). The length of the array is given by
\(
S={\left( N+1\right) }a\). We denote the transverse displacement of element
{\it m} away from the desired control point
by \(
u_{m}\). We will consider boundary conditions that pin the
array at its ends (i.e., at its elements \(
m=0\) and \(
m=N+1\)). There is also a force transverse to the array, of
magnitude \(
f_{m}u_{m}\), acting on element
{\it m} of the chain with \(
f_{m}{\ifmmode>\else$>$\fi}0\), which pushes elements away from the desired
configuration (i.e., \(
u_{m}=0\)) so as to render it unstable in the absence of any
controlling mechanisms. 

 In the simple case where all the couplings have the same value
$\alpha$, the eigenvectors are the normal modes of an extended oscillator,
given by 
\begin{equation}\vcenter{\halign{\strut\hfil#\hfil&#\hfil\cr 
$\displaystyle{e^{{\left( k\right) }}_{m}
=\Rad{{{2}\over{N+1}}}\DoRad \hskip 0.212em 
\sin {{\pi ka}\over{S}}m}$\cr 
}}\end{equation}
An important property of the normal modes is their orthogonality, that
is
\begin{equation}\label{x:orthogonality}\vcenter{\halign{\strut\hfil#\hfil&#\hfil\cr 
$\displaystyle{\sum _{m=1}^{N}e^{{\left( 
j\right) }}_{m}e^{{\left( k
\right) }}_{m}={{2}\over{N+1
}}\sum _{m}\sin {\left( 
{{\pi jam}\over{S}}\right) }\sin 
{\left( {{\pi kam}\over{S}}\right) }
=\delta _{jk}}$\cr 
}}\end{equation}
where \(
\delta _{kj}\) is one when \(
j=k\) and zero otherwise. 

 In this case the matrix {\it A} is
tridiagonal and given by
\begin{equation}\vcenter{\halign{\strut\hfil#\hfil&#\hfil\cr 
$\displaystyle{A_{mn}=
\left\{\matrix{-2\alpha +f&m=n\cr \alpha 
&m=n{\ifmmode\pm\else$\pm$\fi}1\cr 0&\hbox{\rm otherwise}
\cr }\right.}$\cr 
}}\end{equation}
For the corresponding eigenvalues we have
\begin{equation}\vcenter{\halign{\strut\hfil#\hfil&#\hfil\cr 
$\displaystyle{{\left( A{\bf e}^{{\left( k\right) }
}\right) }_{m}={\left( -2\alpha 
+f\right) }e^{{\left( k\right) }
}_{m}+\alpha {\left( e^{{\left( 
k\right) }}_{m+1}+e^{{\left( 
k\right) }}_{m-1}\right) }
}$\cr 
}}\end{equation}
which, with simple trigonometric identities, can be written as
\begin{equation}\vcenter{\halign{\strut\hfil#\hfil&#\hfil\cr 
$\displaystyle{{\left( -2\alpha +f+2\alpha \cos {\left( 
{{\pi ka}\over{S}}\right) }\right) }
e^{{\left( k\right) }}_{m}}$\cr 
}}\end{equation}
When there are no controls, \(
M=A\) and this gives
\begin{equation}\label{x:nocontrol}\vcenter{\halign{\strut\hfil#\hfil&#\hfil\cr 
$\displaystyle{\lambda _{k}=f-2\alpha {\left[ 1-\cos 
{\left( {{\pi ka}\over{S}}\right) }
\right] }}$\cr 
}}\end{equation}

\subsection{ Stability}
 When the system is not controlled, Eqs.~\ref{x:nocontrol} and \ref{x:criterion} 
show the k${}^{th}$
mode of the system is unstable when
\begin{equation}\vcenter{\halign{\strut\hfil#\hfil&#\hfil\cr 
$\displaystyle{f{\ifmmode>\else$>$\fi}f^{{\left( k\right) }}_{%
crit}\equiv 2\alpha {\left( 1-\cos {{%
\pi ka}\over{S}}\right) }\approx {{%
\alpha \pi ^{2}a^{2}k^{2}}\over{%
S^{2}}}}$\cr 
}}\end{equation}
where the last expression holds in the limit of \(
S\gg ak\) or equivalently \(
N\gg k\). This critical value is very close to the one
determining the buckling instability of elastic frames~\cite{naschie90}. 
When {\it f}
is greater than this minimum amount, the solutions given by
Eq.~\ref{x:solution} grow exponentially in
time. 

 The magnitude of the critical force increases with increasing
{\it k}, so the lowest mode is most easily
destabilized. Thus, the shorter the wavelength of the perturbation, the
smaller the amount of stabilizing force needed on each element to keep
the system from going unstable. Conversely, long wavelength
instabilities need stronger stabilizing forces. 

 For simplicity, we focus on the case where the uncontrolled system
has a few unstable modes, say up to \(
k{\ifmmode\leq\else$\leq$\fi}I\). That is, we take \(
f^{{\left( I+1\right) }}_{crit}
{\ifmmode>\else$>$\fi}f{\ifmmode>\else$>$\fi}f^{{\left( I\right) }}_{%
crit}\). We also consider the case of relatively modest
damping so that the higher modes exhibit oscillatory behavior rather
than simple overdamped relaxation. These oscillations pose a challenging
control problem since local displacements can differ from that of the
global behavior of the lower modes to be controlled. Specifically, from
Eqs.~\ref{x:eigenvalues} and \ref{x:nocontrol} the requirement of a 
modest damping constrains its value
to be 
\begin{equation}\vcenter{\halign{\strut\hfil#\hfil&#\hfil\cr 
$\displaystyle{\Gamma ^{2}/4{\ifmmode<\else$<$\fi}f^{{\left( I\right) }
}_{crit}}$\cr 
}}\end{equation}
If we also have \(
\Gamma ^{2}/4{\ifmmode<\else$<$\fi}f^{{\left( I+1\right) }
}_{crit}-f\), then all higher modes are underdamped even with the
destabilizing force.

\subsection{ Control Algorithm}
 To study the effectiveness of different organizations we select a
simple control procedure. Each agent estimates the amplitude of the
unstable modes based on its available information about other agents. It
then pushes with a force proportional to these amplitudes. 

 With this focus on the lowest modes, the agents attempt to devote
their control capabilities only into the modes that requires
stabilization. We assume the agents are aware of the ideal functional
form of the modes of the system so that the estimate can be obtained by
a least squares fit to the available amplitudes, which is a linear
relation. This assumption is relaxed in Sec.~\ref{x:realistic}. 

 Specifically, suppose agent {\it m}
has access to {\it L} values
\(
{\left\lbrace x_{l},v_{l}\right\rbrace }
\), for \(
l=1,\ldots ,L\), where \(
x_{l}\) is a set of agents and \(
v_{l}\) is the average value of the displacements of those
agents. We define \(
e^{{\left( i\right) }}_{x_{l}
}\) as the average value of the eigenvector for those
agents, i.e., \(
{\left( 1/{\left| x_{l}\right| }
\right) }\sum _{k\in x_{l}}
e^{{\left( i\right) }}_{k}
\). Given these values the agent
{\it m} performs a weighted least square
fit to estimate the modes to be controlled. That is, the agent attempts
to distinguish that part of the displacement in unstable modes, needing
control, from that in stable modes which do not need to be controlled.
The number of modes estimated is limited by the smaller of the number of
unstable modes in the uncontrolled system,
{\it I}, and the number of different
values, i.e., {\it L}. The fit is obtained
by minimizing
\begin{equation}\vcenter{\halign{\strut\hfil#\hfil&#\hfil\cr 
$\displaystyle{\delta =\sum _{l=1}^{L}W_{%
lm}{\left( v_{l}-\sum _{i=1}
^{\min {\left( I,L\right) }}
a_{i}e^{{\left( i\right) }}_{%
x_{l}}\right) }^{2}}$\cr 
}}\end{equation}
where the \(
W_{lm}\) are the weights used by the agent and the sum over
{\it i} includes all modes to be estimated.
This can be cast into a standard least squares form by defining
\(
\MthAcnt {v}{\tilde }_{l}=v_{l}\Rad{
W_{lm}}\DoRad \) and \(
\MthAcnt {D}{\tilde }_{li}=e^{{\left( 
i\right) }}_{x_{l}}\Rad{
W_{lm}}\DoRad \) so that \(
\delta ={\left| \MthAcnt {D}{\tilde }{\bf a}-
\MthAcnt {{\bf v}}{\tilde }\right| }^{%
2}\). 

 The most direct method for finding this fit is by setting
\(
{{d\delta }\over{da_{j}}}=0\), which gives
\begin{equation}\vcenter{\halign{\strut\hfil#\hfil&#\hfil\cr 
$\displaystyle{\sum _{l=1}^{L}{\left( 
\sum _{i}\MthAcnt {D}{\tilde }_{%
li}a_{i}-\MthAcnt {v}{\tilde }_{l}
\right) }\MthAcnt {D}{\tilde }_{lj}
=0}$\cr 
}}\end{equation}
for every value of {\it j}, so that the
estimate of the amplitudes is given by the solution to
\begin{equation}\label{x:mode equation}\vcenter{\halign{\strut\hfil#\hfil&#\hfil\cr 
$\displaystyle{\MthAcnt {D}{\tilde }^{T}\MthAcnt {D}{\tilde }
{\bf a}=\MthAcnt {D}{\tilde }^{T}\MthAcnt {{\bf 
v}}{\tilde }}$\cr 
}}\end{equation}
a linear equation for the mode amplitudes. 

 If the positions and weights are chosen so that \(
\sum _{l}W_{lm}e^{{\left( 
i\right) }}_{x_{l}}e^{{\left( 
j\right) }}_{x_{l}}\) is zero unless \(
i=j\), i.e., the terms of the fit are orthogonal, then
\begin{equation}\label{x:mode fits}\vcenter{\halign{\strut\hfil#\hfil&#\hfil\cr 
$\displaystyle{a_{i}={{\sum _{l=1}^{L}
W_{lm}v_{l}e^{{\left( i\right) }
}_{x_{l}}}\over{\sum _{l=1
}^{L}W_{lm}{\left( e^{{\left( 
i\right) }}_{x_{l}}\right) }^{%
2}}}}$\cr 
}}\end{equation}
For example, this holds in the special case where the fit is to just one
mode. 

 More generally, when the terms of the fit are not orthogonal,
Eq.~\ref{x:mode equation} can be used to solve
for the amplitudes {\bf a}. However, this can
give large numerical errors, especially when one of the modes is close
to zero at the location of agent {\it m}.
The amplitude corresponding to such a mode is not well determined, and a
more robust method~\cite{golub83} relies on the
pseudoinverse \(
\MthAcnt {D}{\tilde }^{+}\) of the matrix \(
\MthAcnt {D}{\tilde }\) to give \(
{\bf a}=\MthAcnt {D}{\tilde }^{+}\MthAcnt {{\bf 
v}}{\tilde }\). This will generally set to zero the estimated
amplitudes of modes that are zero at the location of the agent. 

 Given this estimate, the
{\it m${}^{th}$} agent evaluates
the total displacement in the modes to be controlled given by
\(
\sum _{i}a_{i}e^{{\left( 
i\right) }}_{m}\) and then exerts a force proportional to this net
displacement, with the proportionality constant given by
{\it s}. In terms of the control matrix
{\it C}, this force is the value of
\(
{\left( C{\bf u}\right) }_{m}\). 

 To complete the specification of this control algorithm requires
giving the values used and the weights for the fit. These depend on the
particular organizational structures and are described below. 

\subsubsection{ Global}
 In a global organization each agent has access to the state of all
other agents and all agents count equally, i.e., \(
W_{lm}=1\). In terms of the general control discussion given
above, this means \(
L=N\), \(
x_{l}={\left\lbrace l\right\rbrace }\) and \(
v_{l}=u_{l}\). This will generally be done by sending all the
sensor information to a central controller, which then determines the
fit and broadcasts the result to all the agents. Thus all agents use the
same amplitude values and these are exactly the correct values for the
current mode amplitudes. Because of Eq.~\ref{x:orthogonality}, the fits 
are given by Eq.~\ref{x:mode fits} so the control matrix becomes
\begin{equation}\vcenter{\halign{\strut\hfil#\hfil&#\hfil\cr 
$\displaystyle{C_{mn}=s\sum _{i}{{e^{%
{\left( i\right) }}_{m}e^{{\left( 
i\right) }}_{n}}\over{\sum 
_{k=1}^{N}{\left( e^{{\left( 
i\right) }}_{k}\right) }^{%
2}}}=s\sum _{i}e^{{\left( 
i\right) }}_{m}e^{{\left( i
\right) }}_{n}}$\cr 
}}\end{equation}

 In this case, \(
{\bf e}^{{\left( k\right) }}\) is still an eigenvector of the full control matrix
\(
M=A-C\). This is because, with Eq.~ \ref{x:orthogonality},
\begin{equation}\vcenter{\halign{\strut\hfil#\hfil&#\hfil\cr 
$\displaystyle{{\left( C{\bf e}^{{\left( k\right) }
}\right) }_{m}=\sum _{n}
C_{mn}e^{{\left( k\right) }
}_{n}=s\sum _{i}{{e^{%
{\left( i\right) }}_{m}}\over{%
\sum _{l}{\left( e^{{\left( 
i\right) }}_{l}\right) }^{%
2}}}\sum _{n}e^{{\left( 
i\right) }}_{n}e^{{\left( k
\right) }}_{n}=s\delta _{k{\ifmmode\leq\else$\leq$\fi}I}
\hskip 0.212em e^{{\left( k\right) }}_{%
m}}$\cr 
}}\end{equation}
where \(
\delta _{k{\ifmmode\leq\else$\leq$\fi}I}=1\) for \(
k{\ifmmode\leq\else$\leq$\fi}I\) and 0 for larger
{\it k}. That is,
{\it C} multiplies unstable modes by
{\it s}, but gives zero for all other
modes. Thus the eigenvalues of the matrix
{\it M} become 
\begin{equation}\vcenter{\halign{\strut\hfil#\hfil&#\hfil\cr 
$\displaystyle{\lambda _{k}=-f^{{\left( k\right) }
}_{crit}+f-s\delta _{k{\ifmmode\leq\else$\leq$\fi}I}}$\cr 
}}\end{equation}
This implies that the control mechanism only affects the behavior of the
unstable modes. 

 From Eq.~\ref{x:criterion}, the condition
for the stability of the lowest mode, and hence all higher modes as
well, is
\begin{equation}\label{x:smin}\vcenter{\halign{\strut\hfil#\hfil&#\hfil\cr 
$\displaystyle{s{\ifmmode\geq\else$\geq$\fi}f-f^{{\left( 1\right) }}_{%
crit}}$\cr 
}}\end{equation}

\subsubsection{ Local}
 In this case agent {\it m} has only a
single value, namely its own displacement. Thus \(
L=1\), \(
x_{1}={\left\lbrace m\right\rbrace }\) and \(
v_{1}=u_{m}\). With only a single value, the weights are arbitrary,
and we take \(
W_{1m}=1\). In this case the agent only has data to estimate one
mode, namely the lowest one. Eq.~\ref{x:mode fits} gives
\begin{equation}\vcenter{\halign{\strut\hfil#\hfil&#\hfil\cr 
$\displaystyle{a_{1}={{u_{m}}\over{e^{{\left( 
1\right) }}_{m}}}}$\cr 
}}\end{equation}
so that the control force is \(
sa_{1}e^{{\left( 1\right) }}_{%
m}=su_{m}\). Thus the control matrix is proportional to the
identity matrix: \(
C_{mn}=s\delta _{mn}\). 

 The corresponding eigenvalues for the matrix
{\it M} are 
\begin{equation}\label{x:localeigenvalue}\vcenter{\halign{\strut\hfil#\hfil&#\hfil\cr 
$\displaystyle{\lambda _{k}=-f^{{\left( k\right) }
}_{crit}+f-s}$\cr 
}}\end{equation}
and Eq.~\ref{x:criterion} again gives
Eq.~\ref{x:smin} as the stability requirement. In
this case the control also acts (unnecessarily) to further stabilize
higher modes.

\subsubsection{ Hierarchy}
 For hierarchical systems, one must also decide on the appropriate
number of levels, or branching ratio in the hierarchy. We suppose the
weights are \(
W_{lm}=r^{d_{lm}}\) with \(
d_{lm}\) the ultrametric distance between elements
{\it l} and
{\it m}, and
{\it r} characterizes the reduction in
interaction strength that two agents undergo when they are separated by
one further level. 

 Consider now agent {\it m}. In our
hierarchical control organization, each agent receives information from
each of its ancestors, indicating averaged displacement. In this case
\(
L=d+1\), \(
x_{l}\) is the set of the agents situated at ultrametric
distance \(
l-1\) or less and \(
v_{l}\) is the average displacement of all these agents.
These averaged values are given by
\begin{equation}\vcenter{\halign{\strut\hfil#\hfil&#\hfil\cr 
$\displaystyle{v_{l}}$\hfilneg&$\displaystyle{{}={{1}\over{b^{l-1}
}}\sum _{{\left\lbrace i|d_{im}
{\ifmmode\leq\else$\leq$\fi}l-1\right\rbrace }}u_{i}}$\cr 
$\displaystyle{}$\hfilneg&$\displaystyle{{}={{1}\over{b^{l-1}}}
\sum _{i=0}^{b^{l-1}-1}u_{%
\eta {\left( m,d-l+1\right) }+i}}$\cr 
}}\end{equation}
and
\begin{equation}\vcenter{\halign{\strut\hfil#\hfil&#\hfil\cr 
$\displaystyle{x_{l}}$\hfilneg&$\displaystyle{{}={\left\lbrace i|d_{im
}{\ifmmode\leq\else$\leq$\fi}l-1\right\rbrace }}$\cr 
$\displaystyle{}$\hfilneg&$\displaystyle{{}={\left\lbrace \eta {\left( m,d-
l+1\right) },\ldots ,\eta {\left( m,d-l+
1\right) }+b^{l-1}-1\right\rbrace }
}$\cr 
}}\end{equation}
with \(
\eta {\left( m,j\right) }\) given by Eq.~\ref{x:eta(m,j)}.
This information is obtained from the ancestor at level
\(
d-l+1\) of the tree, and hence involves ultrametric distances
up to \(
l-1\). The weight associated with this information is
\(
W_{lm}=r^{l-1}\). 

 In this case, the fit functions with the weights are not
necessarily orthogonal, so instead of Eq.~\ref{x:mode fits}, 
we must use the more general
Eq.~\ref{x:mode equation}. In this case
\(
\MthAcnt {v}{\tilde }_{l}=v_{l}\Rad{
r^{l-1}}\DoRad \), \(
\MthAcnt {D}{\tilde }_{li}=e^{{\left( 
i\right) }}_{x_{l}}\Rad{
r^{l-1}}\DoRad \) and the control force exerted by agent
{\it m} is \(
s\sum _{i}{\left( \MthAcnt {D}{\tilde }^{%
+}\MthAcnt {{\bf v}}{\tilde }\right) }_{%
i}e^{{\left( i\right) }}_{m
}\). What is the contribution from agent
{\it n} to the force exerted by agent
{\it m}? Suppose \(
d_{mn}=k\). Then \(
u_{n}\) appears only in the values \(
v_{l}\) for which \(
l-1{\ifmmode\geq\else$\geq$\fi}k\), and with coefficient \(
1/b^{l-1}\). Thus in \(
\sum _{l}{\left( \MthAcnt {D}{\tilde }^{%
+}\right) }_{il}\MthAcnt {v}{\tilde }_{%
l}\), \(
u_{n}\) enters with coefficient \(
\sum _{l{\ifmmode\geq\else$\geq$\fi}k+1}{\left( \MthAcnt {D}{\tilde }^{%
+}\right) }_{il}\Rad{r^{l-1}
}\DoRad /b^{l-1}\). Hence, the control matrix becomes
\begin{equation}\vcenter{\halign{\strut\hfil#\hfil&#\hfil\cr 
$\displaystyle{C_{mn}=s\sum _{i}\sum 
_{l=d_{mn}+1}^{d+1}{{1}\over{%
b^{l-1}}}{\left( \MthAcnt {D}{\tilde }^{%
+}\right) }_{il}\Rad{r^{l-1}
}\DoRad e^{{\left( i\right) }}_{%
m}}$\cr 
}}\end{equation}
The dependence on node {\it n} enters only
through the ultrametric distance between it and node
{\it m}, thus satisfying the requirement
for a hierarchical control matrix that if node \(
n_{1}\) and node \(
n_{2}\) have the same ultrametric distance to node
{\it m} then \(
C_{mn_{1}}=C_{mn_{2}}\). 

 The resulting control matrix has the same eigenvectors as the local
and global cases for the controlled modes. This means that the stability
criterion is the same as Eq.~\ref{x:smin}. The
remaining, uncontrolled, modes are more complex with this control
organization but do not influence the stability.

\subsubsection{ Multihierarchy}
 The multihierarchical organization is basically the same as the
hierarchical one, except for a change in the definition of the
ultrametric distance. Instead of using a single hierarchy to define the
distance, now a different tree structure is used for each agent. This
means that the averaged values used to define \(
x_{l}\) and \(
v_{l}\) differ from those of the hierarchy, but the general
description remains the same, in that these values are averaged over all
agents at ultrametric distance \(
l-1\) from agent {\it m}, but
now {\em using the ultrametric distance function for agent
{\it m}}. In computing these sets
we ignore any values outside the range of the array. 

 For agent {\it m} we have
\(
x_{l}={\left\lbrace \eta ^{{\left( {\rm mi
n}\right) }}_{m,l-1},\ldots \eta ^{%
{\left( {\rm max}\right) }}_{m,l
-1}\right\rbrace }\). Furthermore
\begin{equation}\vcenter{\halign{\strut\hfil#\hfil&#\hfil\cr 
$\displaystyle{v_{l}={{1}\over{N_{m,l-1}}}
\sum _{i=\eta ^{{\left( {\rm min}\right) }
}_{m,l-1}}^{\eta ^{{\left( {\rm 
max}\right) }}_{m,l-1}}
u_{i}}$\cr 
}}\end{equation}
where we use the definitions of Eqs.~\ref{x:eta(min/max)} and \ref{x:Nmk}. 
The derivation then follows as
for the hierarchy, giving
\begin{equation}\vcenter{\halign{\strut\hfil#\hfil&#\hfil\cr 
$\displaystyle{C_{mn}=s\sum _{i}\sum 
_{l=d_{mn}+1}^{L_{m}}{{%
1}\over{N_{m,l-1}}}{\left( \MthAcnt {D}{\tilde }^{%
+}\right) }_{il}\Rad{r^{l-1}
}\DoRad e^{{\left( i\right) }}_{%
m}}$\cr 
}}\end{equation}
with the ultrametric distance \(
d_{mn}\) being defined with respect to agent
{\it m} as described in
Sec.~\ref{x:multihierarchy}. Different agents can
have differing numbers of values, due to the variation in the number of
levels to reach the edge of the chain. Specifically, \(
L_{m}\) is the smallest value for which both
\(
m-1{\ifmmode\leq\else$\leq$\fi}{\left\lfloor b^{L_{m}-1}/2\right\rfloor }
\) and \(
N+1-m{\ifmmode\leq\else$\leq$\fi}{\left\lfloor b^{L_{m}-1}/2\right\rfloor }
\). This can be expressed as \(
L_{m}={\left\lceil \log_{b}{\left( 2\max {\left( m-1,
N+1-m\right) }\right) }\right\rceil }
+1\). 

 As with the hierarchy, the resulting control matrix has the same
eigenvectors as the local and global cases for the controlled modes.
This means that the stability criterion is the same as
Eq.~\ref{x:smin}. The remaining, uncontrolled,
modes are more complex but do not influence the stability.

\subsection{ Results}
 The effect of different organizations is seen in the forces
required to stabilize the system as a function of the number of
agents. 

 Specifically, we consider how the {``}same material{''}
scales with increasing number of agents. We suppose that when agents are
inactive they don{'}t change the physical properties of the material to be
controlled, so they have no influence on the mass per unit length of the
elastic array. Since the dispersion relation of
Eq.~\ref{x:nocontrol} has the modes, including
the lowest one, depending on {\it N}; i.e.,
\(
\lambda _{k}\approx f-\alpha \pi ^{2}k^{2
}/N^{2}\) it suggests that the appropriate scaling to maintain
the low mode frequencies is to have \(
\alpha \sim N^{2}\). Furthermore, we will consider the case where the
first two modes are unstable, i.e., \(
I=2\), so that \(
f^{{\left( 3\right) }}_{crit}
{\ifmmode>\else$>$\fi}f{\ifmmode>\else$>$\fi}f^{{\left( 2\right) }}_{%
crit}\). In this case, since \(
a\sim 1/N\), {\it f} will be
independent of {\it N.} We will choose
\(
\alpha =1\) for \(
N=8\) and scale $\alpha$ to keep the frequency of the lowest
mode constant. Hence, from Eq.~\ref{x:criterion},
\(
\alpha \approx {{N^{2}}\over{64}}\). With these values we can select \(
f=0.8\) to make the first two modes unstable. To construct
the stability diagram we choose the smallest control matrix that gives
stability, which from Eq.~\ref{x:smin} means
\(
s=f-f^{{\left( 1\right) }}_{crit
}\).  

\begin{figure}
\hspace*{\fill}
\epsfig{file=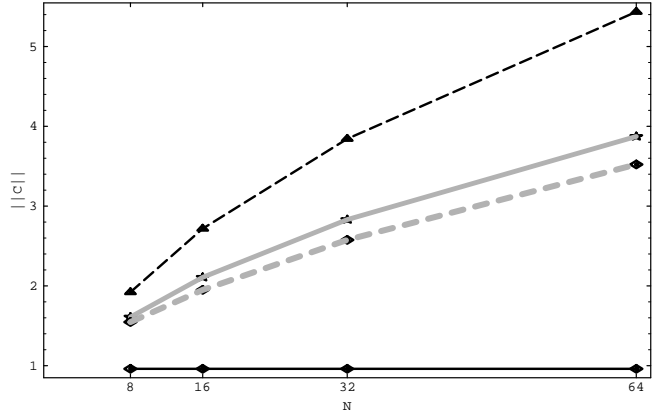,width=3.5in}
\hspace*{\fill}
\caption{\label{x:scaling}Scaling of the norm of the minimal control 
matrix required to achieve stability for various organizations: 
global (solid, black), local (dashed, black), hierarchy (solid, gray) 
and multihierarchy (dashed, gray) as a function of N. 
The hierarchical organizations are chosen with \(r=0.5\) and branching ratio 2.}
\end{figure}

 Fig.~\ref{x:scaling} shows the norm of the
control matrix {\it C} as a function of
{\it N} for these values, as given by the
Frobenius norm~\cite{golub83}, \(
{\left| C\right| }=\Rad{
\sum _{ij}{\left| C_{ij}\right| }^{%
2}}\DoRad \). It shows that the global control can stabilize the
system with the smallest control matrix, as expected since it does not
act (unnecessarily) on the higher modes, which are already stable. The
local control requires the largest matrices. The hierarchical cases are
intermediate, with the multihierarchy being slightly better. This figure
can also be viewed as a stability diagram: for each organization, below
the line the control matrix is too small to stabilize the system, while
above it the system is stable. 

 Notice the difference between the behaviors of the hierarchy and
multihierarchy. Both provide effective ways of giving information to the
agents about dynamics at different length scales, but the multihierarchy
as described above is able to match perturbations at any scale and any
location on the array. It also has the flexibility of combining
information from different parents. 

\begin{figure}
\hspace*{\fill}
\epsfig{file=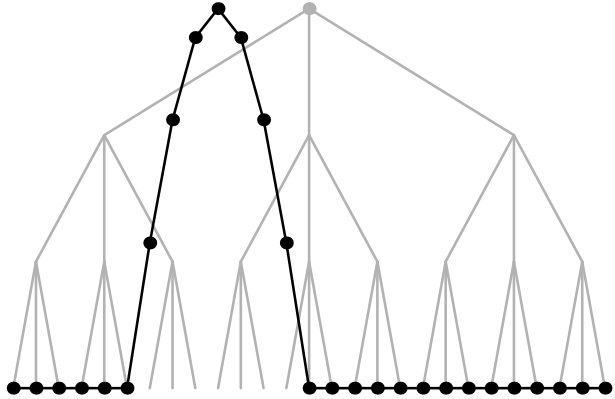,width=3.5in}
\hspace*{\fill}

\caption{\label{x:mismatch}A localized perturbation for a chain with 27 
elements (black) superimposed on the hierarchical organization with 
3 levels and branching ratio 3 (gray). This perturbation cuts across 
a major organizational division of the hierarchy, between agents 9 and 10. 
The perturbation is nonzero only for agents 7 through 13, inclusive.}
\end{figure}

\begin{figure}
\hspace*{\fill}
\epsfig{file=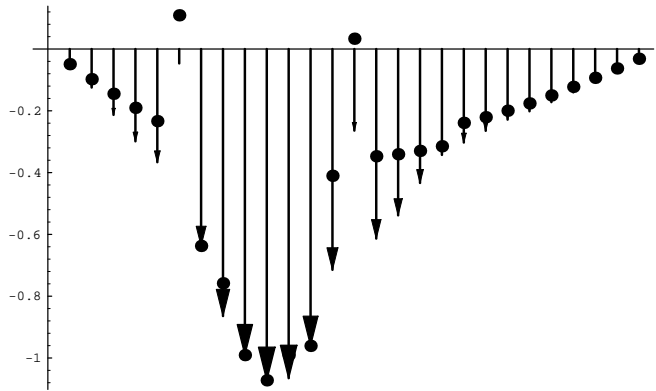,width=3.5in}
\hspace*{\fill}

\caption{\label{x:mismatch forces}Forces acting on each agent when 
the chain is started in the configuration shown in Fig.~\ref{x:mismatch}
 (and with zero velocity) when the maximum of the perturbation is set 
to \(u_{10}=1\). The arrows show the forces produced using the 
multihierarchical control, while the individual points show the forces 
when the hierarchical control is used. In both cases, the agents use 
the maximum amplitude in the lowest mode estimated by their ancestors 
to determine their control response. System parameters are \(\alpha =1\),
 \(f=0.08\) and \(s=1\) so this system is well within the stable regime.}
\end{figure}

 For example, if the agents respond to the maximum amplitude
estimate reported by any ancestors, the multihierarchy will be better
able to respond to perturbations that cross organizational boundaries in
the hierarchical organizations. This is illustrated in
Fig.~\ref{x:mismatch forces}, which depicts the
forces acting on each agent for a localized perturbation shown in
Fig.~\ref{x:mismatch}. Near the maximum of the
perturbation, both organizations respond nearly equally since the local
value of the perturbation gives the largest amplitude estimate. However,
for agents further away, we see the multihierarchy provides a stronger
response since it is more effective at recruiting those agents to help
push against the perturbation. Because the multihierarchy is able to
focus effort at the most appropriate scale {\em and
location}, we can expect it will provide a variety of more
effective control strategies than the hierarchy.

\section{\label{x:realistic} Realistic Smart Matter}
 Traditional control theory and the discussion in the previous
sections assume the existence of good models of the system. In real
systems however, the values of the displacements given by sensors and of
the actuator forces might be imperfect. Thus the agent{'}s knowledge of
the dynamics of the system is only approximate.  

 These differences between theoretical models and actual systems
have many causes. For example, any large scale manufacturing process
leads to artifacts whose functional properties are distributed around a
mean with dispersion $\sigma$, so that the information from sensors and the
forces exerted by actuators will differ from the actual values according
to the dispersion in their properties. This view is appropriate for
devices manufactured in large quantities (e.g., as in many proposed
applications of MEMS) but less so for unique structures or artifacts in
different specific environments (such as control for large structures
such as buildings). 

 Another source of imperfections is due to the {``}smart
dust{''} approach in which control elements are scattered throughout
a structure~\cite{web.mems96}. In such a situation
the local environment of each of the elements won{'}t be constant through
the system thus introducing a random modulation of their
properties. 

 Finally, imperfections or unknown variations in physical
characteristics of system mean the agents fit to an incorrect model of
the dynamics. 

 In all these cases, the entries of the control matrix,
{\it C}, in Eq.~\ref{x:matrix dynamics} would also deviate from their ideal values. In
this section we investigate the consequences for stability of this
randomization of the control matrix, both in general terms and for the
example of the elastic array. 

\subsection{ Stability}
 To investigate the consequences of realistic systems departing from
the idealized models, we consider the average behavior of smart matter
given an ensemble specifying uncertain knowledge about parameters
entering the {\it M} matrix. For the sake
of treating a very general case, we assume as little knowledge about
these parameters as possible. This implies that all matrices can be
considered, and that there is no particular basis for choosing one over
the other. This is the class of the so-called random matrices, in which
all such matrices are equally likely to occur. Matrices in this class
have each entry obtained from a random distribution with a specified
mean and variance. By taking this approach, we can make general
statements about the stability of these systems as the number of
components and degrees of freedoms grows. 

 To do this, we need to determine the average behavior of the
largest eigenvalue of random matrices. The precise value of the
eigenvalue with the largest real part, \(
\lambda _{1}\), of the matrix {\it M
}depends on the particular choice of the entries.
Methodologically, the study of the general behavior of these matrices is
performed by examining the average properties of the class that
satisfies all the known information about them. A class of plausible
matrices is determined by the amount of information one has. The net
results of these considerations is that the full control dynamics can be
written as in Eq~\ref{x:matrix dynamics}, but
with the matrix {\it M} given by the sum of
an ideal part and a random part
\begin{equation}\vcenter{\halign{\strut\hfil#\hfil&#\hfil\cr 
$\displaystyle{M=\MthAcnt {M}{\tilde }+R}$\cr 
}}\end{equation}
both of which are influenced by the choice of control
organization. 

 The eigenvalues of {\it M} are related
to the eigenvalues of \(
\MthAcnt {M}{\tilde }\) and {\it R}. For
example, when the matrices are symmetric or else have non-negative
entries, the largest eigenvalue is real and \(
\lambda _{1}\) is bounded by
\begin{equation}\label{x:bound}\vcenter{\halign{\strut\hfil#\hfil&#\hfil\cr 
$\displaystyle{\MthAcnt {\lambda }{\tilde }_{N}+\lambda ^{%
{\left( R\right) }}_{1}{\ifmmode\leq\else$\leq$\fi}\lambda _{%
1}{\ifmmode\leq\else$\leq$\fi}\MthAcnt {\lambda }{\tilde }_{1}+
\lambda ^{{\left( R\right) }}_{1
}}$\cr 
}}\end{equation}
where \(
\MthAcnt {\lambda }{\tilde }_{N}\) is the smallest eigenvalue of the matrix
\(
\MthAcnt {M}{\tilde }\). Other types of matrices have more complicated
relations concerning the largest eigenvalue but in most cases stability
will be related to the behavior of the eigenvalues of
{\it R}. 

 The eigenvalues of random matrices have been studied for a number
of cases. They show surprising regularities that we can use to discuss
the stability properties of realistic systems. In the case of random
symmetric matrices whose entries have a positive mean value, on average,
the largest eigenvalue is given by~\cite{furedi81}
\begin{equation}\label{x:symmetric}\vcenter{\halign{\strut\hfil#\hfil&#\hfil\cr 
$\displaystyle{\lambda ^{{\left( R\right) }}_{%
1}={\left( N-1\right) }\mu +\sigma ^{%
2}/\mu }$\cr 
}}\end{equation}
 where $\mu$ is the expected value of the entries in
{\it R}, which we assume to be positive.
Moreover, the actual values (as opposed to the average) of
\(
\lambda _{1}\) are normally distributed around this value with
variance \(
2\sigma ^{2}\). When added to \(
\MthAcnt {M}{\tilde }\) this implies that as the size of the system grows,
most such matrices will have positive largest eigenvalue, thus making
the control unstable. The case of asymmetric matrices can also be
treated~\cite{furedi81,juhasz82} to give
\begin{equation}\label{x:non-symmetric}\vcenter{\halign{\strut\hfil#\hfil&#\hfil\cr 
$\displaystyle{\lambda ^{{\left( R\right) }}_{%
1}\sim N\mu }$\cr 
}}\end{equation}
but in general Eq.~\ref{x:bound} will not hold,
making it more difficult to give definite results for this case. 

 From these considerations, we can expect that random matrices are
unstable whenever their size and the magnitude and variation of their
entries are large. In terms of the corresponding organization, these
corresponds to large systems, with many strong interactions among the
agents. This prediction is tempered by the fact that in localized
organizations so many of the matrix elements are zero so that even with
strongly interacting agents they tend to be stable~\cite{hogg89}. 
Which of these effects is dominant is
determined by the organizational structure.

\subsection{ Control Organizations}
 We now describe how the predictions concerning random matrices
apply to different control organizations. 

\subsubsection{ Global}
 Since a global control fills the full matrix, the imperfections
make the matrix random. Assuming for simplicity that the matrix
{\it R} has entries all chosen from the
same probability distribution with mean $\mu$ and variance
\(
\sigma ^{2}\), the full matrix {\it M}
will have a mean given by
\begin{equation}\vcenter{\halign{\strut\hfil#\hfil&#\hfil\cr 
$\displaystyle{\MthAcnt {M}{\bar }_{ij}=\MthAcnt {M}{\tilde }_{%
ij}+\mu }$\cr 
}}\end{equation}
and each element has variance \(
\sigma ^{2}\).  

 Since the largest eigenvalue of the ideal matrix is negative due to
its stability under controls, as {\it N}
gets large the upper bounds of Eq.~\ref{x:bound}
gets positive. This happens even in the case of moderate values of
{\it N}, as the dispersion of the
imperfection properties will make the right hand side positive.

\subsubsection{ Local}
 A local control matrix is diagonal, so symmetry is quite natural in
contrast to the global case. For a diagonal matrix
{\it R}, the largest eigenvalue is given by
the largest diagonal component. Since theses are chosen from a random
distribution with mean $\mu$, and variance \(
\sigma ^{2}\), the relevant eigenvalues originate in the extrema of
the distribution, and they are given by the extreme value distribution
function. If the eigenvalues are normally distributed, the extrema will
scale with {\it N} as~\cite{woodroofe75} \(
\lambda ^{{\left( R\right) }}_{1
}\sim \sigma \Rad{2\ln N}\DoRad \), implying higher stability than 
the global case due
to the slow growth with {\it N} of
\(
\lambda ^{{\left( R\right) }}_{1
}\). This scaling is independent of the value of $\mu$
provided it does not grow rapidly with
{\it N}. The extrema of other distributions
also grow quite slowly compared to
{\it N}~\cite{woodroofe75}, showing that the increased stability of
the local architecture as compared with the global is a general
property, provided the matrix elements come from distributions with the
same mean and variance.

\subsubsection{ Hierarchies and Multihierarchies}
 Suppose now that the system has a hierarchical structure of depth
{\it d} and average branching ratio,
{\it b}. In this case, the strength of the
interaction between any two agents decreases with their ultrametric
distance. Specifically, the interaction strength, i.e., the size of the
element in the control matrix, will be taken to be \(
r^{h}\), with {\it h} the number
of hierarchy levels that separate the two agents. 

 Detailed treatment of the properties of random matrices with
hierarchical structure remains an open problem. Nevertheless, a good
qualitative insight is obtained by ignoring this detailed structure
except for its effect on the mean and variance of the distribution from
which the matrix elements are selected. The size of the matrix is given
by \(
N=b^{d}\) and the mean of the non-diagonal terms is
\begin{equation}\label{x:mu}\vcenter{\halign{\strut\hfil#\hfil&#\hfil\cr 
$\displaystyle{\mu ={{\sum _{h=1}^{d}
b^{h-1}{\left( b-1\right) }r^{h}
}\over{\sum _{h=1}^{d}b^{h-1
}{\left( b-1\right) }}}={{%
{\left( rb\right) }^{d}-1}\over{b^{%
d}-1}}\hskip 0.265em {{r{\left( b-1
\right) }}\over{rb-1}}}$\cr 
}}\end{equation}
The nature of the organization depends on whether the total interaction
is dominated by neighboring agents or distant ones. The first case
amounts to restricting {\it r} to be
\(
r{\ifmmode<\else$<$\fi}1/b\). This choice makes the decreasing interaction
strength between agents overwhelm the increase in their number as higher
levels in the hierarchy are considered. In this situation when
{\it N} is large
Eq.~\ref{x:mu} becomes \(
\mu ={{r{\left( b-1\right) }}
\over{{\left( 1-rb\right) }N}}\), which implies that as
{\it N} grows the mean goes to zero,
leading to a stable system because of Eq.~\ref{x:non-symmetric}. 
However, if fluctuations in the coupling strength
were taken into account, some of the matrices could still become
unstable. 

\begin{figure}
\hspace*{\fill}
\epsfig{file=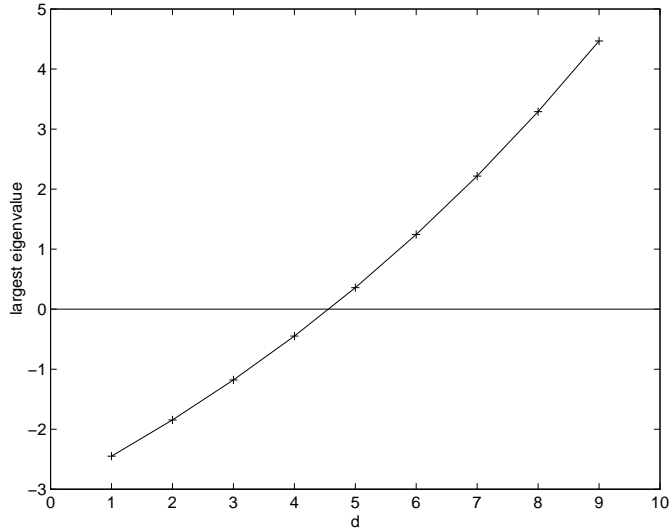,width=3.5in}
\hspace*{\fill}

\caption{\label{x:r055}Plot showing the growth of the largest eigenvalue 
of a hierarchical random matrix with branching ratio \(b=2\) and \(r=0.55\) 
as a function of \(d=\log_{2}n\). The diagonal elements were chosen to be 
\(-3\). The points are the theoretically predicted values based on 
Eq. \ref{x:mu} and lie very close to the computed eigenvalues shown by the solid line.}
\end{figure}

 In the second case, \(
r{\ifmmode>\else$>$\fi}1/b\). This implies that in the large
{\it N} limit the average becomes
\begin{equation}\vcenter{\halign{\strut\hfil#\hfil&#\hfil\cr 
$\displaystyle{\mu =r^{d}{{r{\left( b-1\right) }
}\over{rb-1}}={{r{\left( b-1\right) }
}\over{rb-1}}N^{{{\ln r}\over{\ln 
b}}}}$\cr 
}}\end{equation}
Note that since \(
-1{\ifmmode<\else$<$\fi}{{\ln r}\over{\ln b}}{\ifmmode<\else$<$\fi}
0\), \(
\mu \) goes to zero as the size of the matrix grows, but in
slower fashion than the case above. Nevertheless, this subtle difference
in convergence to zero amounts to a qualitative difference in the
stability of the system. To see this, notice that Eq. \ref{x:non-symmetric} 
implies that the largest eigenvalue of a random
matrix grows as \(
\mu N\), which increases with
{\it N} for this value of \(
\mu \), thus leading to instability when the system gets
large enough. The growth in largest eigenvalue with the size of the
system is exhibited in Fig.~\ref{x:r055} for a
particular choice of parameters. Notice that the system becomes unstable
for \(
d{\ifmmode\geq\else$\geq$\fi}5\). Given these results we see that the size of the
matrix for which this instability takes place is much larger than the
one in the absence of a hierarchy of interactions. 

 Therefore,  the largest eigenvalue will be similar to the behavior
of the global case with a renormalization of the  $\mu$ given by
Eq.~\ref{x:mu}. Thus, a hierarchical control
system is more stable than the global one but less so than the local
case, for given interaction strength. 

 The multihierarchy organization matrix has a structure similar to
the hierarchical case, but different in the sense that the entries are
centered around the diagonal. This greater symmetry does not affect
their values and therefore one gets the same qualitative behavior as for
the hierarchy. A more detailed analysis of the different structures
would be required to distinguish the behavior of these two
organizations.

\subsection{ Example: The Imperfect Elastic Array}
 Elucidating the factors that dominate the behavior of random
matrices for different organizations requires looking at particular
instances. To see how imperfections affect the control of the elastic
array, we consider the three sources of imperfections that arise in such
systems and assume the errors they introduce are multiplicative. The
sensors at the {\it j${}^{th}$}
location gives its reading of the local value as \(
u^{{\rm meas}}_{j}\). Because of its imperfect nature, this reading will
differ from the actual value, \(
u_{j}\), by an amount \(
u^{{\rm meas}}_{j}=u_{j}{\left( 
1+r_{j}\right) }\), where \(
r_{j}\) represents the random imperfection of the sensor.
Similarly for the actuators, the actual force, \(
f^{{\rm actual}}_{i}\) will differ from the nominal one, \(
f_{i}\), by
\begin{equation}\vcenter{\halign{\strut\hfil#\hfil&#\hfil\cr 
$\displaystyle{f^{{\rm actual}}_{i}=f_{i}{\left( 
1+e_{i}\right) }}$\cr 
}}\end{equation}
where \(
e_{i}\) represents the random imperfection of the actuator.
The last source of imperfection has to do with the unknown variation in
physical characteristics of the system to be controlled. This translates
in having control matrix entries differ from the ideal values,
\(
C_{ij}\) by \(
C^{{\rm computed}}_{ij}=C_{ij}{\left( 
1+\rho _{ij}\right) }\). 

 Combining all these factors gives
\begin{equation}\vcenter{\halign{\strut\hfil#\hfil&#\hfil\cr 
$\displaystyle{f^{{\rm actual}}_{i}=-s\sum 
C_{ij}{\left( 1+r_{j}\right) }
{\left( 1+e_{i}\right) }{\left( 
1+\rho _{ij}\right) }u_{j}\equiv 
-s\sum _{j}C^{{\rm actual}}_{%
ij}u_{j}}$\cr 
}}\end{equation}
 which we can write as \(
C^{{\rm actual}}_{ij}=C_{ij}{\left( 
1+\epsilon _{ij}\right) }\) where \(
\epsilon _{ij}\) includes contributions from all the sources of
errors. If we suppose that the deviations from the nominal values are
independent random values, but there is no systematic bias, then the
\(
\epsilon _{ij}\) can be viewed as random variables with zero mean and
some standard deviation we denote by $\sigma$. The introduction of errors
can make the matrix \(
C^{{\rm actual}}_{ij}\) asymmetric, even though the ideal case is symmetric.
For the full control dynamics governed by the matrix
{\it M}, the value of \(
M_{jk}\) will vary randomly with an average value
\begin{equation}\vcenter{\halign{\strut\hfil#\hfil&#\hfil\cr 
$\displaystyle{A_{jk}+C_{jk}}$\cr 
}}\end{equation}
and standard deviation \(
\sigma C_{jk}\). 

\begin{figure}
\hspace*{\fill}
\epsfig{file=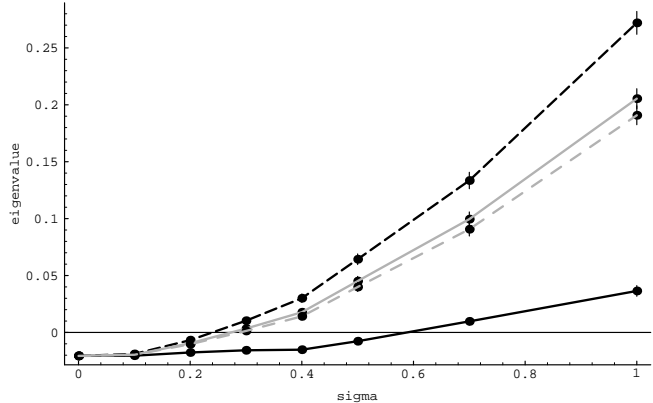,width=3.5in}
\hspace*{\fill}

\caption{\label{x:noise}Effect of noise on the expected value of the largest 
eigenvalue, for \(N=8\), \(\alpha =1\), \(f=0.8\) and controls with \(s=0.7\) 
as a function of $\sigma$ for asymmetric multiplicative noise with zero mean 
and standard deviation $\sigma$. The organizations are global (solid, black), 
local (dashed, black), hierarchy (solid, gray) and multihierarchy (dashed, gray). 
For the hierarchical cases, we have \(r=0.5\) and a branching ratio of two. 
The error bars give the standard error estimate of the mean, based on 1000 samples. 
Eigenvalues above zero indicate situations where the noise makes the system unstable,
on average.}
\end{figure}

 Fig.~\ref{x:noise} shows the behavior of the
expected value of the largest eigenvalue for a particular case with
different organizations. The local control is most sensitive to the
noise. This is contrary to what one might expect from analyses based on
the connectivity of the matrices and is due to larger matrix elements,
shown in Fig.~\ref{x:scaling}, being more
important than the degree of connectivity in the matrices. 

 These comparisons used the same set of random matrices for each
organization, thus allowing a detailed pairwise comparison~\cite{press86} 
of their relative performance independent of
the variation due to different choices of the noise matrices. This shows
that the difference in the behavior of the organizations is
statistically much more significant than appears from the error bars in
the figure. 

 Fig.~\ref{x:noise} shows the behavior of the
average of \(
\lambda _{1}\). Another measure of the consequence of imperfection
is the fraction of cases that are unstable at each value of
\(
\sigma \). This shows the same qualitative behavior of the
organizations typically becoming unstable in the same order as shown for
the average value. 

 This example exhibits the phenomenon of instability brought about
by the system size, the interconnectivity of the control organization
and the variance in the imperfections.

\section{ The Effect of Delays}
 In our discussion, we assumed that the agents were able to act
instantaneously on the physical system. This is appropriate when the
control elements (sensors, actuators and computations) operate rapidly
compared to the relevant physical time scales of the unstable system. On
the other hand, delays could arise from a variety of causes. For
instance, there could be communication delays, in which case larger
organizational structures, e.g., global, would have more delay than
local ones. There could also be different communication delays for
different agents as for example depending on their ultrametric distance
in a hierarchy. However, while passing information through multiple
levels of managers in a hierarchical organization may introduce
additional delays in the information, for controlling physical devices
this may not be much of a problem: the information aggregated over
larger spatial scales corresponds to lower frequency behaviors which are
less sensitive to delays. 

 On the other hand, if the main delay is in activation of the
actuators, e.g., due to their inertia, then the delays could be more
significant and fairly similar among the organization choices. A final
possibility is that delays arise from the computation time required to
determine the response to the given sensor inputs. This is not likely to
be a problem for the simple control mechanisms discussed here but could
be more significant if more complex strategies are used, especially if
the agents include expectations of the behavior of other agents in their
evaluations. 

 The behavior of the system with delays in the control can be
analyzed most readily by assuming that all delays in the information are
the same so Eq.~\ref{x:matrix dynamics} becomes
\begin{equation}\vcenter{\halign{\strut\hfil#\hfil&#\hfil\cr 
$\displaystyle{{{d^{2}{\bf u}{\left( t\right) }
}\over{dt^{2}}}=-\Gamma {\bf \MthAcnt {u}{\dot }
}{\left( t\right) }+A{\bf u}{\left( 
t\right) }-C{\bf u}{\left( t-\tau \right) }
}$\cr 
}}\end{equation}
where $\tau$ is the delay. That is, the control decisions are made based
on a delayed value for the displacements. In this case the stability
analysis is similar to that given above but results in a more
complicated criterion. Specifically, suppose \(
{\bf u}{\left( t\right) }={\bf u}_{0}
e^{\Lambda t}\) then we have
\begin{equation}\vcenter{\halign{\strut\hfil#\hfil&#\hfil\cr 
$\displaystyle{{\left( \Lambda ^{2}I+\Gamma \Lambda I-A+C
e^{-\Lambda \tau }\right) }{\bf u}_{%
0}=0}$\cr 
}}\end{equation}
This has a solution only when the matrix is singular, i.e., when
\begin{equation}\label{x:delay criterion}\vcenter{\halign{\strut\hfil#\hfil&#\hfil\cr 
$\displaystyle{\det {\left( \Lambda ^{2}I+\Gamma \Lambda 
I-A+Ce^{-\Lambda \tau }\right) }=0}$\cr 
}}\end{equation}
This is a generalization of the eigenvalue criterion for the case with
no delays, and is a so-called exponential-polynomial~\cite{bellman63}. 
Unlike the dispersion relation in the
undelayed case, this generally has an infinite number of complex
solutions. Stability corresponds to having the real parts of
{\em all} the solutions being nonpositive. 

 As an example, consider the case of the chain for the lowest mode
\(
{\bf e}^{{\left( 1\right) }}\) and assume the control matrix is such that this is an
eigenvector of both {\it A} and
{\it C}, with eigenvalues \(
\lambda ^{{\left( A\right) }}\) and \(
\lambda ^{{\left( C\right) }}\), respectively. By comparison, the dynamics with no
delays is governed by \(
M=A-C\) with eigenvalue \(
\lambda =\lambda ^{{\left( A\right) }}
-\lambda ^{{\left( C\right) }}\). In that case the criterion becomes
\begin{equation}\vcenter{\halign{\strut\hfil#\hfil&#\hfil\cr 
$\displaystyle{\Lambda ^{2}+\Gamma \Lambda -\lambda ^{{\left( 
A\right) }}+\lambda ^{{\left( C\right) }
}e^{-\Lambda \tau }=0}$\cr 
}}\end{equation}

 For a simple analysis, consider the effect of a small delay, so
\(
e^{\Lambda \tau }\approx 1-\Lambda \tau \) near the largest solutions. Then we have
\begin{equation}\vcenter{\halign{\strut\hfil#\hfil&#\hfil\cr 
$\displaystyle{\Lambda ^{2}+{\left( \Gamma -\lambda ^{%
{\left( C\right) }}\tau \right) }
\Lambda -\lambda =0}$\cr 
}}\end{equation}
This effectively amounts to a decrease in the damping, but is otherwise
the same as the condition leading to Eq.~\ref{x:eigenvalues} 
for the case without delays. 

 When the delays prevent achieving stability, more sophisticated
control algorithms will be required. For example, instead of a least
squares fit to delayed values, the agents could use some averaging of
different estimates of the size of the first mode, or attempt to
extrapolate to the current state of the system. These sophisticated
control strategies can be helpful, but some caution is required since
they can also introduce additional instabilities~\cite{kephart89PhysRevA}. 

\subsection{ Example: Elastic Array with Control Delays}
 In the chain example, \(
\lambda ^{{\left( C\right) }}=s\) is positive so to first order, 
a delay decreases the
damping a bit but does not change the stability properties, provided
this decrease does not make the overall damping negative, i.e., as long
as \(
s\tau {\ifmmode<\else$<$\fi}\Gamma \). While this argument is only valid for small delays,
it shows how delays in the information available to the agents can lead
to instability. Note that we require \(
s{\ifmmode\geq\else$\geq$\fi}\lambda ^{{\left( A\right) }}\) to have stability 
even with no delays in the control.
Thus these requirements restrict \(
\Gamma /\tau {\ifmmode>\else$>$\fi}s{\ifmmode\geq\else$\geq$\fi}\lambda ^{{\left( A\right) }
}\) and suggest there is no stable control beyond
\(
\tau {\ifmmode>\else$>$\fi}\Gamma /\lambda ^{{\left( A\right) }
}\). This analysis for a single mode can also apply to
other modes of the system. However, higher modes, i.e., those with
smaller or even negative (in the case of modes that are stable even
without control) values of \(
\lambda ^{{\left( A\right) }}\) can tolerate larger delays than the lowest mode.
Hence stability is determined by the behavior of the lowest mode, as is
the situation when there are no delays. A more complete analysis of the
stability~\cite{bellman63} gives only a small
correction to this criterion as long as the delays are small. An
interesting observation from this criterion is that at least some
damping is essential for the system to be stable with even the smallest
delay. 

\begin{figure}
\hspace*{\fill}
\epsfig{file=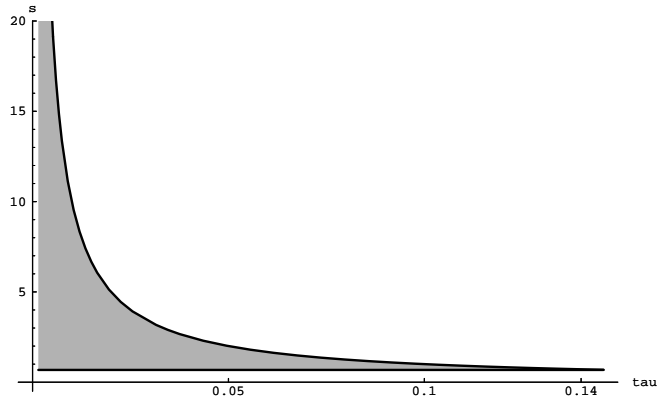,width=3.5in}
\hspace*{\fill}

\caption{\label{x:delay stability}The gray region shows the stable combinations 
of control force, given by {\it s}, and delay $\tau$, for \(N=8\), \(\alpha =1\) 
and \(f=0.8\) with \(\Gamma =0.1\). Stability requires the control to be at least 
\(s{\ifmmode\geq\else$\geq$\fi}f-f^{{\left( 1\right) }}_{crit}=0.679\) 
even with no delay. 
The upper limit is close to \(s{\ifmmode\leq\else$\leq$\fi}\Gamma /\tau \). 
For \(\tau {\ifmmode>\else$>$\fi}0.147\) there is no stable control.}
\end{figure}

 An example is show in Fig.~\ref{x:delay stability}. 
The dynamical equations with delays can also be solved
directly~\cite{bellman63,virk85,marsaglia89} to
exhibit the behavior of the system when the control is delayed.

\section{ Explosive Transient Growths}
 Maintaining stability is an important goal, but there remains the
possibility of large transient growth even when system is stable. That
is, the displacements could become large at intermediate time scales
before eventually damping out. This means that for controlling these
systems, stability is necessary but may not be sufficient to guarantee
that small perturbations never produce large amplitudes which could
couple to nonlinearities and damage the system. 

 This problem arises when the matrix
{\it M} governing the system has some
eigenvectors that are nearly parallel. For symmetric matrices, where the
eigenvectors are always orthogonal, this problem never occurs. For
instance, many physical systems, such as the elastic array we have
considered, have physical forces that are symmetric. This leads to a
symmetric matrix {\it A} provided the
masses of the components are equal, as we assumed in the case of the
elastic array. More general unequal masses, or the addition of an
asymmetric random control matrix due to imperfections could lead to
nearly parallel eigenvectors in some control organizations. Moreover,
there are important systems, such as fluid motion~\cite{trefethen93}, 
where the physical interactions
themselves can give rise to asymmetric matrices with nearly parallel
eigenvectors. 

\begin{figure}
\hspace*{\fill}
\epsfig{file=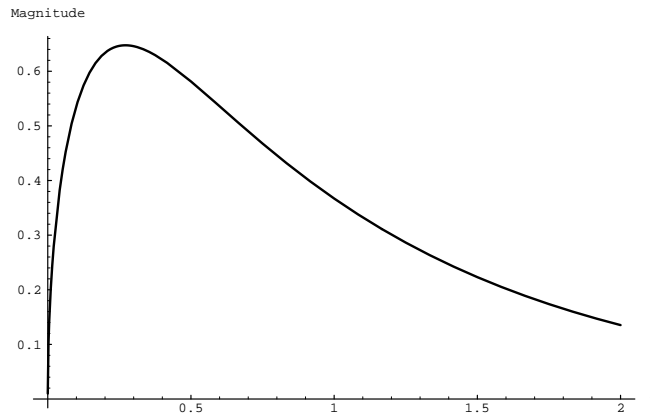,width=3.5in}
\hspace*{\fill}

\caption{\label{x:trans}Example of a giant transient for the two dimensional 
example with \(\epsilon =0.01\). Although the system is asymptotically stable, 
its magnitude becomes many times larger than the initial 
perturbation before eventually decaying.}
\end{figure}

 To illustrate the consequence of nearly parallel eigenvectors on
the transient behavior of a mechanical system with a stable fixed point
consider a two dimensional case where the eigenvectors are
\(
v_{1}={\left( 1,0\right) },\hskip 0.265em 
v_{2}={\left( 1,\epsilon \right) }\), with eigenvalues \(
\lambda _{1}=-1\hskip 0.265em \hbox{\rm and}\hskip 0.265em 
\lambda _{2}=-5\). The dynamics are determined by \(
v{\left( t\right) }=av_{1}e^{\lambda _{%
1}t}+bv_{2}e^{\lambda _{2}t}
\). Thus, for an initial condition where these two
vectors nearly cancel, such as \(
a=1\hskip 0.265em \hbox{\rm and}\hskip 0.265em b=-1\), the evolution becomes \(
v{\left( t\right) }={\left( e^{-t}
-e^{-5t},\hskip 0.212em -\epsilon e^{-5t}\right) }
\) with magnitude \(
e^{-t}\Rad{1-2e^{-6t}+e^{-8t}{\left( 
1+\epsilon ^{2}\right) }}\DoRad \). An example of the resulting dynamics is shown in
Fig.~\ref{x:trans}. 

 The possibility of this large transient growth is not commonly
appreciated in analyses of system stability, and gives another criterion
to check for in designing control organizations. For our example of
multiplicative noise in the elastic array, the local control, which is
diagonal, will not give rise to any asymmetries. However the global
control, and also to some extent, the hierarchical organizations, can be
sensitive to errors in this way. 

\begin{figure}
\hspace*{\fill}
\epsfig{file=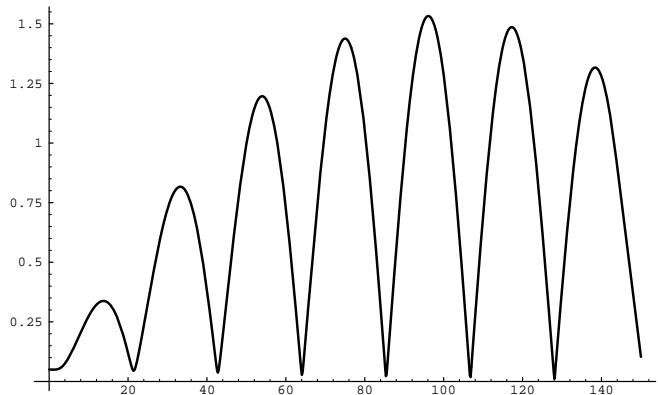,width=3.5in}
\hspace*{\fill}

\caption{\label{x:transient}Example of transient growth possible with asymmetric noise. 
In this case, the magnitude of the displacement vector is shown on the vertical axis 
with \(N=8\), \(\alpha =1\) and \(f=0.8\) with \(\Gamma =0.005\) with global control 
of \(s=0.7\). The horizontal axis depicts time. Initially the system was at rest with 
a displacement vector of magnitude 0.05. Subsequently, the displacement vector reaches 
a magnitude of about 1.5, about 30 times larger than the initial state. 
For larger times, the oscillations continue with gradually decreasing amplitude 
showing that the system is indeed stable.}
\end{figure}

 An example of the elastic array with asymmetric multiplicative
noise giving a large transient growth is shown in
Fig.~\ref{x:transient}. This system is stable,
i.e., the oscillations eventually approach zero. However, the initial
perturbation is greatly amplified before this eventual decrease, thus
providing a situation where the system could couple to undesirable
nonlinear behaviors, or overwhelm the maximum forces available from the
actuators, even though the stability criterion is satisfied.

\section{ Discussion}
 In this paper, we studied the behavior of control organizations for
unstable physical systems. We have shown how a hierarchical organization
is a reasonable compromise between rapid local responses with simple
communication and the use of global knowledge. This holds not only in
ideal situations but also when imperfections and delays are present in
the system. This work illustrates the consequence of the strong physical
embedding of the computational agents. 

 We also introduced a new control organization, the multihierarchy,
and showed it is somewhat better than a hierarchy in achieving
stability, while having in addition a position invariant response which
allows for the control of disturbances at appropriate scale and
location. 

 An important characteristic of smart matter is the distinction
between the use of physical information vs. social or computational
information: when the control consists in minor adjustments and the
communication is fast, agents can respond based on the physical states
of the system. In other cases, e.g., when there are long delays or very
strong actuation forces, it may be important for the agents to respond
based on expectations for what the other agents are likely to do. This
leads to the possibility of more complex dynamics, as has been seen in
computational ecosystems~\cite{kephart89PhysRevA}.
Equally interesting there is regime between these extremes where the
dynamics will significantly couple the networks of computational
interactions and the physical substrate thus giving rise to novel
behaviors. 

 Although the behaviors of distributed controls were illustrated in
the context of a one-dimensional elastic array, the results apply to a
wider range of situations. First because this oscillatory model
describes a variety of physical systems, these results hold for two and
three dimensional materials, as well as giving some insight into the
control of nonlinear systems. Thus from knowledge of the physical
properties of the system to be controlled one is able to characterize
its controllability with distributed computation. 

 Second, we have treated imperfections independently for each agent.
As the number of agents increases and they are closer together, this
assumption may fail. For example, a defect at a certain location may
affect all the agents close to it giving rise to short range
correlations in the entries of the random matrix describing the control.
Longer range correlations may be due to systematic mistakes in the
agents{'} models of the system. While random matrices with correlations
can be studied and show qualitatively similar behaviors, in our case
short range correlations will have no effect on the stability. This is
because stability is determined by the behavior long wavelengths
excitations of the physical system, and these low modes will not be
affected by short range correlations. 

 Third, this methodology applies to other performance metrics. These
include a variety of goals for smart matter: stability; small transient
growth; robustness against imperfections in physical model, sensors,
actuators and delays; simplicity of programming control; low
communication and power use; and minimal cost of the devices. Generally,
we will desire some combination of these goals. Since these goals can
conflict, a major challenge for control design is how to manage them.
This points to the need for a simple framework to express trade-offs
among these goals. For large distributed systems, one appropriate
mechanism is given by computational markets~\cite{waldspurger92,clearwater96}, 
with the advantage of
simplicity, robustness and scalability. 

 We close by pointing out some open issues that remain to be
studied. One is the question of intermittent failures of actuators and
sensor noise. Unlike the static imperfections we treated, they require
stochastic controls~\cite{astrom70}. This raises
the question of how different organizations respond to ongoing noise, or
sudden changes (e.g., a stuck actuator). The latter example amounts to a
sudden switch from one static imperfect system to another, to which our
results should apply. Another issue is the reaction time scales of
different organizations, and whether there are more appropriate control
mechanisms to deal with the large transient growth that we
discussed. 

 Other questions that remain to be addressed include dynamically
adjusting organization based on changes. For example, as components are
assembled into larger structures the relative performance of different
organizations may change, leading to the evolution of different
structures~\cite{huberman95}. Similarly some parts
of the structure may be more active than others, requiring locally
different organizational structures. In these cases agents may have less
overall information than we assumed in our study (e.g., knowledge of
nominal global modes and their location within the overall structure)
leading to further possibilities to explore. These include different,
and dynamically changing, organizational structures as the agents learn
about the nature of the system they work with. Another question is to
what extent the physical propagating modes in the underlying substrate
can be used as a communication channel to convey properties of the
environment, thereby limiting the need for fabricating additional
wires. 

 Controlling smart matter brings together domains of physics and
distributed computing. As a result, the physical characteristics of the
system to be controlled strongly influence the effectiveness of the
computational organizations. Smart matter thus provides a fertile ground
for the interplay between dynamics in Euclidean space and processing in
abstract spaces.

\section{ Acknowledgement}
 We have profited from many useful comments by Mark Yim. This work
was partially supported by ONR Contract No.
N00014{--}94-C{--}0096

\end{document}